# Dynamic causal modelling of mitigated epidemiological outcomes

Karl J. Friston[1], Guillaume Flandin[1], Adeel Razi[2]

[1]*The Wellcome Centre for Human Neuroimaging, University College London, UK*
[2]*Turner Institute for Brain and Mental Health & Monash Biomedical Imaging, Monash University, Clayton, Australia*

**E-mails**: *Karl Friston, k.friston@ucl.ac.uk; Guillaume Flandin, g.flandin@ucl.ac.uk; Adeel Razi, adeel.razi@monash.edu*

## Abstract

This technical report describes the rationale and technical details for the dynamic causal modelling of mitigated epidemiological outcomes based upon a variety of timeseries data. It details the structure of the underlying convolution or generative model (at the time of writing on 6-Nov-20). This report is intended for use as a reference that accompanies the predictions in following dashboard: https://www.fil.ion.ucl.ac.uk/spm/covid-19/dashboard

**Key words**: *coronavirus; epidemiology; compartmental models; dynamic causal modelling; variational; Bayesian*

## Contents

# Introduction

Since the introduction of dynamic causal modelling for quantitative prediction of the current coronavirus epidemic (Friston et al., 2020a; Friston et al., 2020b), the structure of the model has been progressively optimised as new data became available. In this technical report, we describe the current structure and provide illustrative predictions of various outcomes at the time of writing (i.e., 6-Nov-2020).

Dynamic causal modelling stands apart from most modelling in epidemiology by predicting *mitigated* outcomes—and quantifying the uncertainty associated with those outcomes. This stands in contrast to the majority of quantitative epidemiological modelling used for forecasting, which considers *unmitigated* outcomes. In other words, the forecasts or projections of the sort most commonly seen in the media (and offered as a basis of policy-making by groups such as the SPI-M[1]) try to predict what could happen on the basis of current trajectories. As a rule of thumb, these predictions are usually over the next few weeks—and rest upon fitting curves to the recent trajectory of various data. In contrast, dynamic causal modelling considers not what could happen, but what is most likely to happen. This mandates a generative model of interventions that mitigate viral transmission, such as social distancing, lockdown, testing and tracing, *etc*. In turn, this requires a detailed consideration of how various sorts of data are generated. For example, fluctuations in testing capacity and sampling bias due to people self-selecting when symptomatic. The advantage of this kind of modelling is that any data generated by the model can be used to inform the model parameters that underwrite fluctuations in latent states, such as the prevalence of infection. Latent states refer to those states of the population that cannot be estimated directly and have to be inferred from observable data.

Dynamic causal modelling focuses not on worst-case scenarios but on the most likely outcomes, given concurrent predictions of viral transmission, responses in terms of behavioural interventions and changes in the way that the epidemic is measured (e.g., confirmed cases, death rates, hospital admissions, testing capacity, *etc*). Crucially, dynamic causal modelling brings two things to the table. The first is the use of variational procedures to assess the quality of—or evidence for—any given model. This means that the model adapts to the available data; in the sense that the best model is taken to be the model with the greatest evidence, given the current data. As time goes on, the complexity of the model increases, in a way that is necessary to explain the data accurately. Technically, log evidence (a.k.a. marginal likelihood) is accuracy minus complexity—and both are a function of the data (Penny, 2012).

The second advantage of dynamic causal modelling is a proper incorporation of uncertainty in the estimation of conditional dependencies. In other words, it allows for the fact that uncertainty about one parameter affects uncertainty about another. This means dynamic causal models generally have a large number of parameters, such that the conditional uncertainty about all the parameters is handled together. This furnishes a model that is usually very expressive and may appear over-parameterised. However, by optimising the prior probability density over the model parameters, one can optimise the complexity (c.f., the effective number of parameters), using Bayesian model selection (Friston et al., 2018; Friston and Penny, 2011; Hoeting et al., 1999). Note that the ability to pursue this form of structure learning rests on being able to estimate the model evidence or marginal likelihood, which is one of the primary *raisons d'être* for the variational procedures used in dynamic causal modelling (Beal, 2003; Friston et al., 2007; Winn and Bishop, 2005).

---

[1] https://www.gov.uk/government/groups/scientific-pandemic-influenza-subgroup-on-modelling

These potential advantages can be leveraged to model a large variety of data types to fit a fairly expressive model of the current epidemic and, implicitly, produce posterior predictive densities over measurable outcomes. In other words, parameterising behavioural responses—such as social distancing—as a function of latent states, enables the model to guess how we will respond in the future, with an appropriate uncertainty. This is the basis of the predictions of mitigated responses mentioned above.

The remainder of this report provides a brief description of current predictions using 10 sorts of data. These posterior predictive densities are based upon the current implementation of DCM for COVID-19 described in the appendix and detailed in the accompanying annotated MATLAB code (see software note).

## Dynamic causal modelling

The convolution, generative or forward model of epidemic data (here, from the United Kingdom) is based upon four factors, each of which corresponds to a distinct kind of latent state: each with a number of distinct levels. One of these factors (the *infection* factor) can be thought of as a conventional epidemiological model. The remaining factors are concerned with population fluxes and fluctuating contacts between people, who may or may not be affected, and the clinical progression of the infection that depends on—but is separate from—the infection factor. This allows for both symptomatic and asymptomatic clinical corollaries of infection. The final factor concerns testing. This is a key part of the generative model because it generates the data generally considered to be informative about the course of the epidemic.

The particular DCM described in the appendix is based upon the original model of a single region (Friston et al., 2020a). It has subsequently been extended to deal with viral spread both within and between communities (Friston et al., 2020b). This was necessary to explain secondary waves[2]. A further difference between early applications of dynamic causal modelling in this setting and current applications is the use of multimodal data. The example in Figure 1 uses 10 sources of data (see appendix for details):

- confirmed cases based upon PCR testing, as reported by specimen date

---

[2] https://www.fil.ion.ucl.ac.uk/spm/covid-19/TR5_Second_Wave.pdf

- daily deaths within 28 days of testing positive for COVID-19, reported by date of death
- critical care unit occupancy as measured by the number of patients requiring mechanical ventilation
- the number of PCR tests performed each day
- the number of people infected, based upon unbiased community surveys using PCR tests
- the percent of people who are seropositive, based upon unbiased community surveys using antibody tests
- the number of people reporting symptoms, as estimated by the COVID symptom tracker
- estimates of the reproduction ratio, issued by the government
- mobility, as measured by Department of Transport estimates of car use
- location, as estimated by Google's mobility data; specifically, the relative probability of being in the workplace

Note that some of these so-called data would be treated as estimates in conventional modelling, for example, the reproduction ratio. However, dynamic causal modelling treats these estimates as data features because they are based upon historical data. In other words, dynamic causal modelling generates the underlying reproduction ratio directly from latent states, such as the rate of change of prevalence of infection. This means it can then predict estimates based upon legacy data, under the assumption that there are random effects that accompany these conventional estimators.

Figure 1 shows the data (black dots) and predictions in terms of posterior expectations (blue lines) and associated 90% credible intervals (shaded areas). Here, we consider the first eight outcomes listed above. Note that the period over which data is available—and the time between observations—varies with different data types. However, because the generative model operates in continuous time from the beginning of the outbreak into the future, all data points can be used. Here, we see that the first and secondary waves of confirmed cases show a marked asymmetry, with a much larger number of confirmed cases in the secondary wave. This is largely a reflection of the number of tests performed, as evidenced by the antisymmetric profile of daily deaths (averaged over seven days), which are predicted to peak at around 200 deaths per day[3] in early November 2020 (specifically, November 8). This pattern is reflected in the number of patients requiring mechanical ventilation, the estimated prevalence of infection from community surveys and the symptoms reported using the COVID symptom tracker.

The model also shows a decline in seropositivity from about 7% to 5% at the time of writing, which is predicted to increase again over the next few weeks. The reproduction ratio started at over 2 and fell, during the first lockdown, to below one, dipping to a minimum of about 0.7 over the summer. After this, it rose to about 1.5 and fell below 1 in October. Note that the conventional estimates of the reproduction ratio—relative to the posterior expectation from DCM—are a slight overestimate. Crucially, the black dots in this figure (upper and lower confidence intervals, based upon consensus from the SPI-M) have been *shifted two weeks backwards* in time from their date of reporting. This is what is meant above by historical or retrospective estimates. In other words, the estimates of the reproduction ratio pertain to states of affairs a few weeks ago. At the time of writing, this was particularly relevant because a national lockdown had just been announced with the aim of getting the reproduction ratio below one. According to this analysis, it was already below one at the time of the announcement[4].

---

[3]  Note that this prediction pertains to death by date, not by date reported, which has exceeded 400 on at least one day at the time of writing.

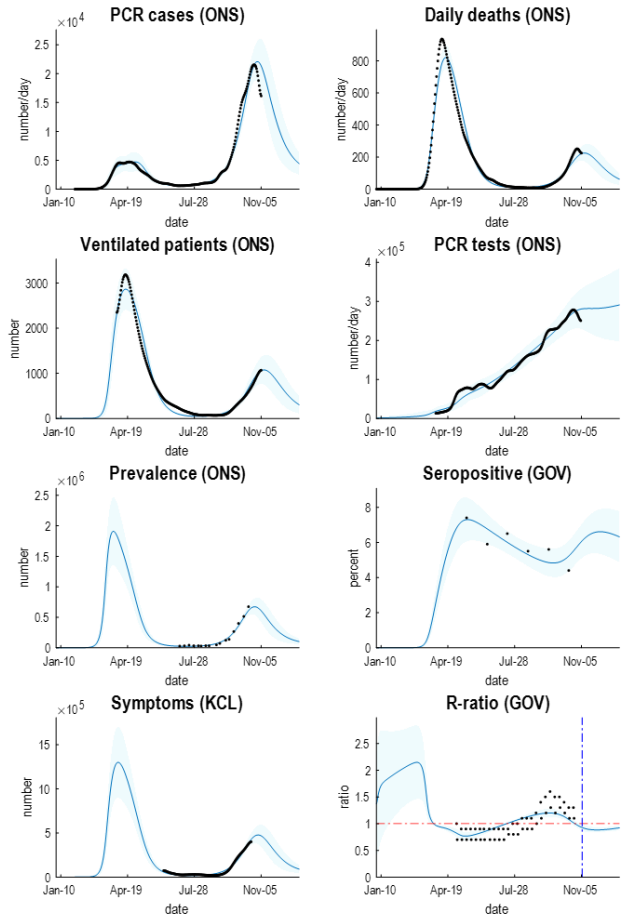

**Figure 1:** *posterior predictions of various outcome modalities, ranging from confirmed cases through to the reproduction ratio*

Figure 2 shows the equivalent results for mobility and location based upon Department of Transport and Google mobility data. It suggests that the national lockdown in spring reduced our contact rates to about 25% of pre-COVID levels; after which they rose again slowly until the resurgence of infections at the onset of the secondary wave. In Figure 2, 100% refers to the pre-COVID mobility.

---

4  https://www.gov.uk/government/news/prime-minister-announces-new-national-restrictions

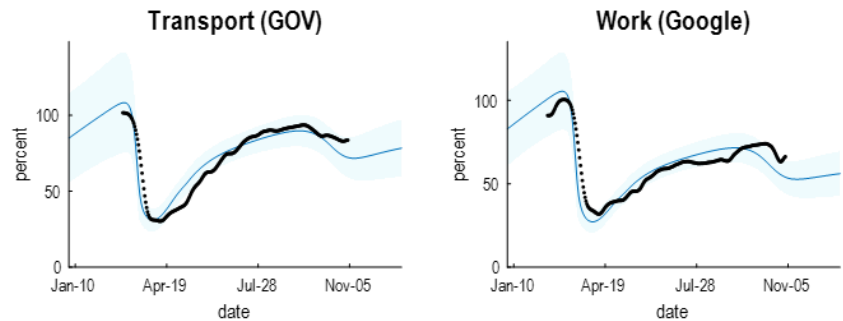

**Figure 2:** *posterior predictions for mobility*

Uncertainty quantification is an inherent aspect of the data assimilation afforded by dynamic causal modelling. In other words, dynamic causal models are convolution models that use Bayesian or variational approaches to assimilate data—and the attending uncertainty. In this particular dynamic causal model, all the uncertainty resides in the model parameters, such as various rate constants and probabilities (see Table 1 and the appendix). This uncertainty is then propagated through to time-dependent latent states and, ultimately, the outcomes (denoted by the shaded confidence intervals in the figures above).

Having said this, variational procedures are notoriously overconfident. They underestimate the uncertainty because of the way they handle conditional dependencies under mean field approximations to the posterior density (MacKay, 2003). To compensate for this, the confidence intervals in the above figures have been inflated by multiplying the posterior standard deviation by a factor of eight. Furthermore, these confidence intervals do not incorporate uncertainty about the structure or form of the model itself. In other words, although the model has been optimised over the past months to maximise model evidence, there is no guarantee that this is the best model, which it will almost certainly not be.

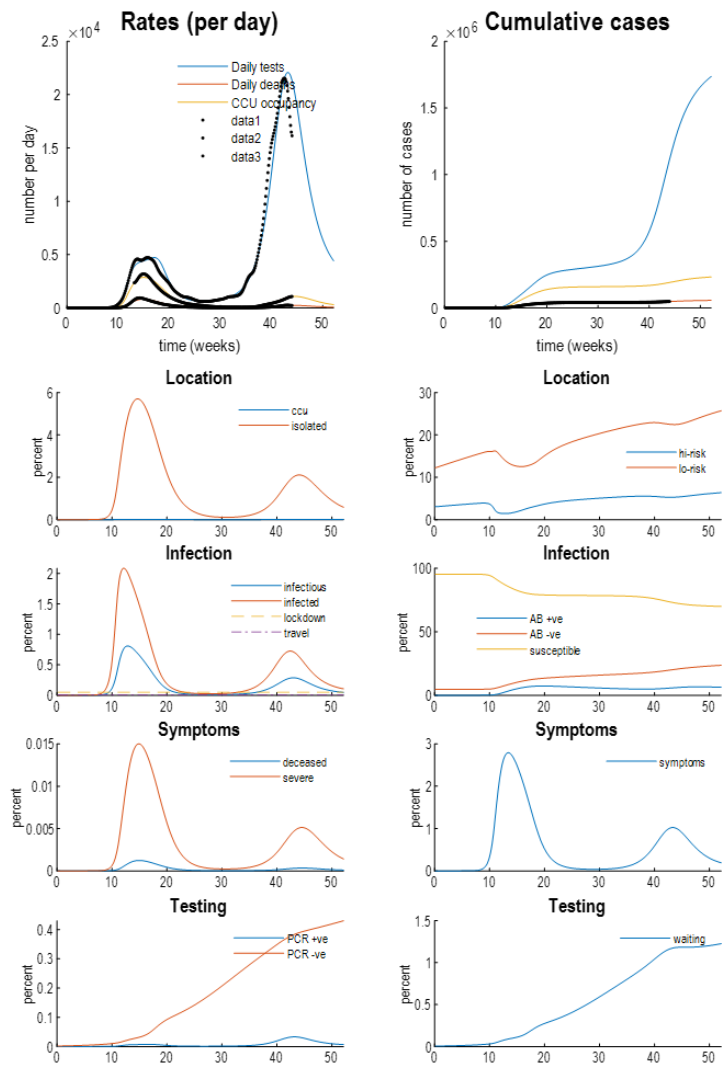

**Figure 3:** *posterior predictions and underlying latent states*

**Figure 3** shows the underlying latent states generating the predictions in Figures 1 and 2. The upper two panels show some outcomes from the previous figures (black dots): daily confirmed cases using PCR testing, daily deaths, and critical care occupancy. The upper left panel shows the rates, while the upper right panel shows the cumulative totals. The remaining panels detail the fluctuations in the latent states of the four factors. Each factor has two panels, showing each of the accompanying levels. For clarity, some levels have been omitted because the probabilities of being in any level—of any given factor—sum to one. For a more detailed explanation of what these latent states mean—and how they can be interpreted—please see the appendix.

The infection panel has been equipped with prior thresholds for restrictions on contacts within (lockdown) and between (travel) an *effective* or active population. It is these thresholds that produce the periodic expression of secondary and subsequent waves following the initial outbreak. One key thing to note here is that, in this example, about 70% of the population remains susceptible to future infection (after January 2021). The remaining population are at low risk of coming into contact with the virus or have acquired an effective immunity; irrespective of whether they are seropositive or seronegative.

**Table 1:** *parameters of the dynamic causal model*

| Number | Name | Description | Prior mean | Precision | Lower (95%) | Upper (95%) | Posterior mean | Lower (95%) | Upper (95%) |
|---|---|---|---|---|---|---|---|---|---|
| 1 | $N$ | population size (M) | 66.65 | Inf | 66.65 | 66.65 | 66.65 | 66.65 | 66.65 |
| 2 | $n$ | initial cases | 1 | 0.25 | 0.037 | 26.8 | 1.04 | 0.41 | 2.6 |
| 3 | $r$ | pre-existing immunity | 0.1 | 16 | 0.066 | 0.15 | 0.047 | 0.032 | 0.069 |
| 4 | $o$ | initially exposed | 0.1 | 16 | 0.066 | 0.15 | 0.14 | 0.13 | 0.15 |
| 5 | $out$ | P(leaving home) | 0.3 | 64 | 0.24 | 0.36 | 0.27 | 0.23 | 0.31 |
| 6 | $sde$ | threshold: distancing | 0.05 | 64 | 0.040 | 0.061 | 0.047 | 0.044 | 0.049 |
| 7 | $qua$ | threshold: quarantine | 0.005 | 64 | 0.0040 | 0.0061 | 0.0095 | 0.0083 | 0.010 |
| 8 | $exp$ | P(leaving area) | 0.0005 | 16 | 0.0003 | 0.0007 | 0.0010 | 0.0009 | 0.0011 |
| 9 | $cap$ | CCU beds per person | 0.00032 | 16 | 0.00021 | 0.00048 | 0.00057 | 0.00043 | 0.00076 |
| 10 | $s$ | distancing sensitivity | 2 | 64 | 1.6 | 2.4 | 2.4 | 2.2 | 2.7 |
| 11 | $u$ | quarantine sensitivity | 6 | 64 | 4.8 | 7.3 | 7.32 | 5.9 | 8.9 |
| 12 | $c$ | mechanical sensitivity | 1 | 64 | 0.81 | 1.2 | 0.79 | 0.65 | 0.96 |
| 13 | $nin$ | contacts: home | 2 | 64 | 1.6 | 2.4 | 1.0 | 0.88 | 1.3 |
| 14 | $nou$ | contacts: work | 64 | 64 | 52 | 78.6 | 81 | 70 | 94 |
| 15 | $trm$ | transmission (early) | 0.5 | 64 | 0.40 | 0.61 | 0.53 | 0.46 | 0.61 |
| 16 | $trn$ | transmission (late) | 0.5 | 64 | 0.40 | 0.61 | 0.27 | 0.23 | 0.31 |
| 17 | $tin$ | infected period (days) | 5 | 1024 | 4.7 | 5.2 | 4.9 | 4.7 | 5.1 |
| 18 | $tcn$ | infectious period (days) | 4 | 1024 | 3.7 | 4.2 | 3.4 | 3.3 | 3.6 |
| 19 | $tim$ | loss of immunity (days) | 256 | 64 | 208 | 314 | 222 | 208 | 237 |

| 20 | *res* | seronegative proportion | 0.4 | 64 | 0.32 | 0.49 | 0.45 | 0.42 | 0.49 |
|---|---|---|---|---|---|---|---|---|---|
| 21 | *tic* | incubation period (days) | 4 | 1024 | 3.7 | 4.2 | 3.0 | 2.8 | 3.1 |
| 22 | *tsy* | symptomatic period (days) | 6 | 1024 | 5.6 | 6.3 | 6.7 | 6.5 | 7.0 |
| 23 | *trd* | ARDS period (days) | 11 | 1024 | 10.4 | 11.5 | 8.8 | 8.5 | 9.1 |
| 24 | *sev* | P(ARDS\|symptoms): early | 0.005 | 1024 | 0.0047 | 0.0052 | 0.0047 | 0.0044 | 0.0049 |
| 25 | *lat* | P(ARDS\|symptoms): late | 0.005 | 0.25 | 0.00018 | 0.13 | 0.0035 | 0.0030 | 0.0041 |
| 26 | *fat* | P(fatality\|ARDS): early | 0.5 | 1024 | 0.47 | 0.52 | 0.44 | 0.42 | 0.46 |
| 27 | *sur* | P(fatality\|ARDS): late | 0.5 | 0.25 | 0.018 | 13.4 | 0.31 | 0.25 | 0.38 |
| 28 | *ttt* | FTTI efficacy | 0.036 | 1024 | 0.034 | 0.037 | 0.037 | 0.035 | 0.039 |
| 29 | *tes* | testing: bias (early) | 1 | 0.25 | 0.037 | 26.8 | 0.22 | 0.017 | 2.9 |
| 30 | *tts* | testing: bias (late) | 4 | 0.25 | 0.14 | 107 | 5.2 | 4.8 | 5.7 |
| 31 | *del* | test delay (days) | 4 | 1024 | 3.7 | 4.2 | 3.9 | 3.7 | 4.1 |
| 32 | *ont* | symptom demand | 0.01 | 0.25 | 0.00037 | 0.26 | 0.0074 | 0.0045 | 0.012 |
| 33 | *fnr* | false-negative rate | 0.2 | 1024 | 0.18 | 0.21 | 0.20 | 0.19 | 0.21 |
| 34 | *fpr* | false-positive rate | 0.002 | 1024 | 0.0019 | 0.0021 | 0.0020 | 0.0019 | 0.0021 |
| 35 | *lin* | testing: capacity | 0.01 | 0.25 | 0.00037 | 0.26 | 0.0065 | 0.0059 | 0.0071 |
| 36 | *rat* | testing: constant | 48 | 0.25 | 1.7 | 1288 | 44 | 41 | 48 |
| 37 | *ons* | testing: onset | 200 | 0.25 | 7.4 | 5367 | 203 | 194 | 213 |

**Sources**:
https://royalsociety.org/-/media/policy/projects/set-c/set-covid-19-R-estimates.pdf
https://arxiv.org/abs/2006.01283

*Table 1 lists the parameters of this model, their priors and their posteriors based upon the data above. The free parameters are listed in the second column and their role in shaping the epidemiological dynamics is described in the appendix.*

Table 1 provides a brief description of the parameters, their prior densities and the posterior density afforded by fitting the data in the figures above. The prior precision corresponds to the inverse variance of the log transformed priors. Although the scale parameters are implemented as probabilities or rates, they are estimated as log parameters. The prior means and ranges were based upon the above sources—and have been progressively optimised with successive versions of the model using Bayesian model reduction. Note that some parameters have narrow (informative) priors, while others are relatively uninformed. The upper and lower ranges of the prior and posterior confidence intervals contain 90% of the probability mass. Note that these are probabilistic ranges, and the posterior estimates can easily exceed these bounds, if the data calls for it.

## Predictive validity

Dynamic causal modelling is generally used to test hypotheses about the causal structure that generates data. This rests exclusively on Bayesian model comparison, where each hypothesis or model is scored using a (a variational bound) on model evidence. This enables one to find the best explanation for the data at hand that has the greatest predictive validity. This follows because cross-validation accuracy goes hand-in-hand with model evidence: in other words, maximising model evidence precludes over fitting by minimising complexity—and ensures generalisation to new data (Hochreiter and Schmidhuber, 1997; MacKay, 2003; Penny, 2012). This is particularly prescient for epidemiological modelling because 'new data' pertains to the future, which means generalisation corresponds to predictive validity. Using models for forecasting that have not been subject to appropriate Bayesian model selection will have poor predictive validity because they overfit, if too complex, or underfit, if not sufficiently expressive. An example is provided in Figure 4[5].

---

[5]  This—and the subsequent example in Figure 6— were added just prior to submission.

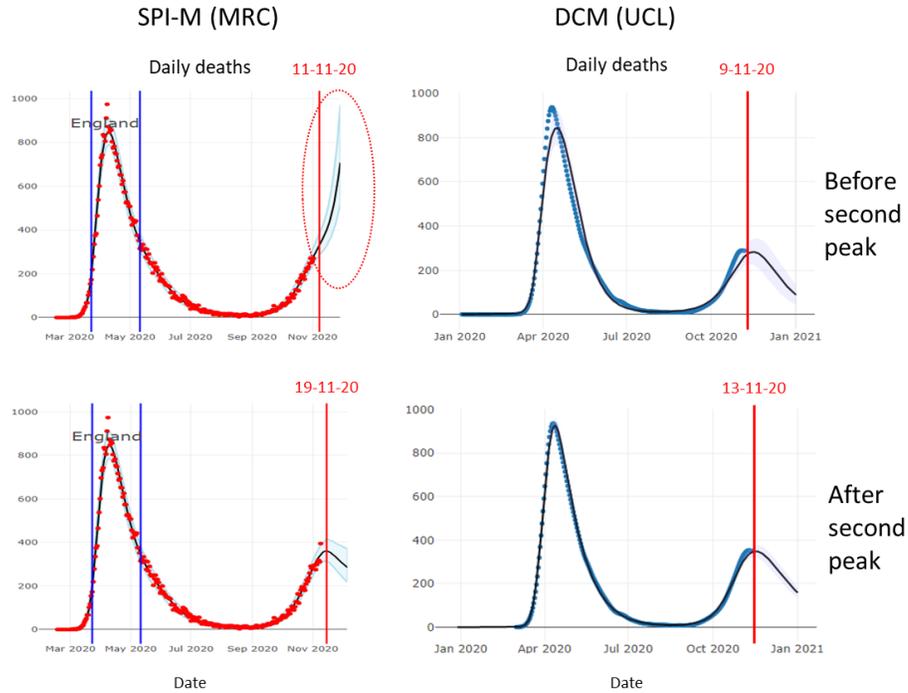

**Figure 4:** *predicting the second peak*

Figure 4 illustrates a failure of nowcasting and forecasting with epidemiological (transmission) models (Birrell et al., 2020) that have not been optimised using variational model comparison. The left panels show the forecasts from the MRC Biostatistics Unit at the University of Cambridge shortly before and after the peak of a second surge. These projections were taken as screen grabs from the dashboard[6] on appropriate days. The right panels show the equivalent predictions using dynamic causal models[7] that have been optimized in terms of model evidence (a.k.a., marginal likelihood). The first MRC forecast on 11-Nov-20 predicted that the number of deaths each day would rise exponentially and "is likely to be between 380 and 610 on the 21st of November". In fact, death rates peaked on 9 November at 398 (seven day average, evaluated on 20-Nov-20). This was predicted by the equivalent dynamic causal modelling. Crucially, the dynamic causal modelling predictions were consistent before and after the peak. Conversely, the MRC predictions had no predictive validity, forecasting opposite trends before and after the peak (indicated with the red ellipse). Furthermore, dynamic causal modelling predicted this peak before an upsurge in confirmed cases, albeit with a smaller amplitude and three weeks earlier (see Figure 5).

---

[6]  https://www.mrc-bsu.cam.ac.uk/tackling-covid-19/nowcasting-and-forecasting-of-covid-19/

[7]  Taken from the dashboard at https://www.fil.ion.ucl.ac.uk/spm/covid-19/dashboard/

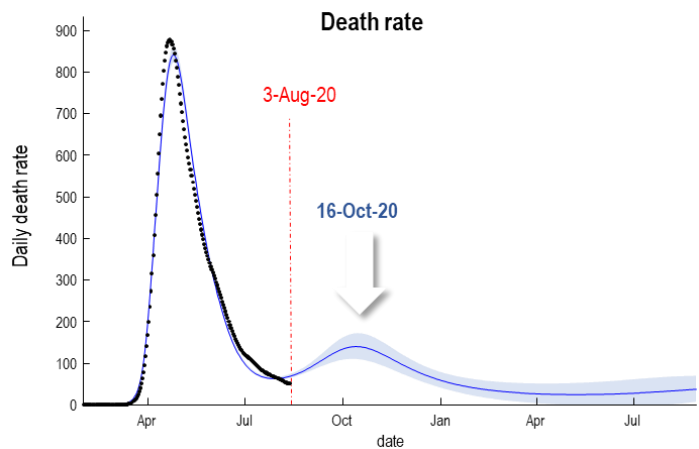

**Figure 5:** *Dynamic causal modelling predictions of a second wave before any increase in confirmed cases in early August. This figure is taken from an internal report on the causes of second waves[8].*

An important application of nowcasting is to estimate the reproduction ratio (or R-number). As noted above this is often used as a point of reference for evaluating when various mitigations should be considered. However, when estimated using models that have not been optimised estimates of things like the reproduction ratio can, in principle, become inaccurate and biased. An example is shown in Figure 6. The left panel shows the reproduction ratio in terms of an expected value (black line) and credible intervals based from the MRC Biostatistics Unit at the University of Cambridge (for London). The right panel shows the upper and lower intervals (blue dots) based upon a consensus of several modelling groups that constitute the SPI-M (for the United Kingdom).

In both instances, the reproduction ratio was estimated to be above one at the time the number of new infections peaked in the United Kingdom. This is mathematically impossible because the R-number should be exactly one at the time of peak incidence. Conversely, the estimates based on dynamic causal modelling (see Figure 6 and appendix) suggest that the reproduction ratio fell below one about 3 weeks before the peak in death rates (black line in the right panel). The dynamic causal modelling estimates have a similar amplitude to the MRC and consensus (SPI-M) estimates; however, the latter appear to lag the former by two weeks. In short, retrospective estimates that are used to motivate various time-sensitive nonpharmacological interventions [9] —and assess their relative impact—may not be apt for guiding time-sensitive decisions.

---

[8]  https://www.fil.ion.ucl.ac.uk/spm/covid-19/TR5_Second_Wave.pdf

[9]  For example, the Consensus Statement on COVID-19, Date: 14th October 2020' states "The number of daily deaths is now in line with the levels in the Reasonable Worst Case and is almost certain to exceed this within the next two weeks. Were the number of new infections to fall in the very near future, this exceedance of the reasonable worst-case scenario might only continue for three to four weeks, but if R remains above 1 then the epidemic will further diverge from the planning scenario."

https://assets.publishing.service.gov.uk/government/uploads/system/uploads/attachment_data/file/931162/S0808_SAGE62_201014_SPI-M-O_Consensus_Statement.pdf

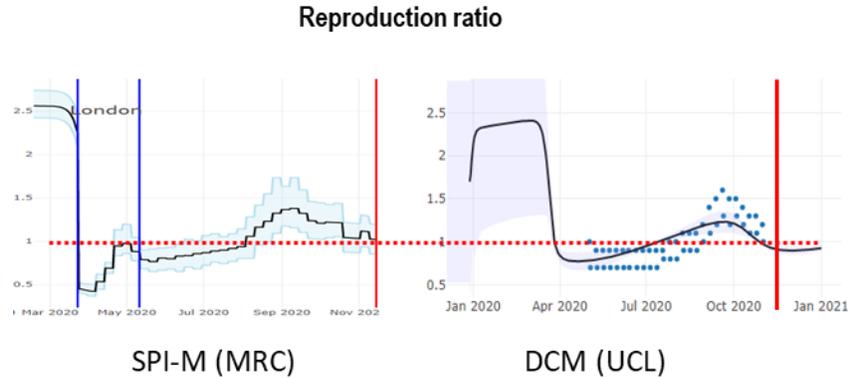

**Figure 6:** *estimates of the reproduction ratio from the MRC Biostatistics Unit (left panel) and dynamic causal modelling (right panel). The black lines correspond to the best (posterior) expectations and the shaded area correspond to credible intervals. The blue dots superimposed over the dynamic causal modelling estimates report the ranges of consensus values reported by the government[10]. These have been plotted two weeks before the end date of the reporting period for each pair.*

## Conclusion

In dynamic causal modelling, everything is optimised with respect to the marginal likelihood or evidence for a model, as scored by a variational free energy or evidence bound. This has the important consequence that the best model is a function of the data at hand. In turn, this means that the best model at the beginning of the epidemic is not the best model halfway through. As more data becomes available, the model needs to become more expressive (or complex) in order to provide an accurate account of these data. The complexity depends upon how tight the priors are over the model's free parameters. There is an optimal complexity that, in conjunction with the accuracy of fit, subtends model evidence.

This optimal complexity is identified using Bayesian model comparison. In other words, the model is defined in terms of which parameters are allowed to vary and which are, *a priori*, more constrained. This model optimisation is itself an adaptive and ongoing process that can, in principle, continue as long as data keeps arriving. As noted above, although Bayesian model comparison (in the form of Bayesian model reduction) has been used throughout the epidemic, there is no guarantee that the basic form of the model—or its coarse graining—is necessarily the best. This would depend upon an exhaustive search of the model space, which is a difficult problem. A problem that may be addressed when the epidemic enters its endemic phase.

---

# Appendix

This appendix describes the form of the generative model. The precise details of the implicit likelihood model and priors can be found in the annotated MATLAB scripts that generated the figures in this report. Details about the standard variational inversion of this model can be found in (Friston et al., 2020a). The trajectory of the epidemic has called for an expressive DCM that can fit a large family of trajectories in different outcome modalities. At first glance, this expressivity may be confused with over-parameterisation. However, the effective degrees of freedom—or effective number of parameters—depend sensitively upon prior probability densities that themselves have been optimised using Bayesian model reduction. In other words, some parameters have uninformative priors and can be considered free parameters, while others have relatively tight or informative priors. The effective number of parameters corresponds to model complexity; namely, Kullback-Leibler divergence between the posterior and the prior. This effectively counts the number of parameters that are used to explain the data.

Currently, the dynamic causal modelling (DCM) has about 40 parameters that parameterise 400 differential or update equations. This may sound like a large number; however, these update equations inherit from fairly common-sense assumptions about transitions among latent states. These structural assumptions, in combination with the prior densities, constitute the assumptions made by the model. At no point do we assume any particular parameter is known: every model parameter is equipped with a greater or lesser posterior uncertainty that is bounded by the priors in Table 1.

The basic form of the DCM is built upon a Master equation (Seifert, 2012) that describes the discrete updates of the probability over the latent states of the model $p \in \mathbb{R}^{5 \times 5 \times 4 \times 4}$, day by day:

$$\vec{p}_{\tau+1} = P \cdot \vec{p}_{\tau}$$
$$\vec{p} = vec(p) \tag{1}$$

This equation can also be expressed as a set of ordinary differential equations by noting the following equivalence:

$$\dot{\vec{p}} = J(p)\vec{p}$$
$$\vec{p}_{\tau+1} = e^{J} \cdot \vec{p}_{\tau} \Rightarrow P = e^{J} \tag{2}$$

Here, $J$ plays the role of the Jacobian of the density dynamics and would have the continuous differential equation form used in most conventional modelling. However, we will stick to the discrete form using the Master equation, based upon a large probability transition matrix $P$. This matrix can be factorized into transitions among the states of each of the four factors (denoted by superscripts). In what follows, we will use $P_{\circ \bullet \bullet \bullet}$ to denote the probability transitions within the first factor states, $P_{\bullet \circ \bullet \bullet}$ denotes probability transitions within the second factor, and so on. Here, the dot stands in for of all levels of the factor corresponding to the order of the index. Similarly, $p_{\circ\bullet\bullet\bullet}$ (note the lower-case $p$) represents the marginal distribution having averaged over the factors denoted by the black dots. For example, $p_{1\bullet\bullet\bullet}$ is the probability of being in the first state of the location factor, marginalised over all other factors. The matrix factorization is as follows:

$$P = P^{(1)} \cdot P^{(2)} \cdot P^{(3)} \cdot P^{(4)}$$

$$P^{(1)} = I_4 \otimes \begin{bmatrix} I_2 \otimes P_{\circ\bullet,1\bullet} & & & \\ & I_2 \otimes P_{\circ\bullet,2\bullet} & & \\ & & I_2 \otimes P_{\circ\bullet,3\bullet} & \\ & & & I_2 \otimes P_{\circ\bullet,4\bullet} \end{bmatrix} \tag{3}$$

$$P^{(2)} = \ldots$$

This means that we can build the model by considering transitions among states of each factor in turn and then compose these factor-specific probability transition matrices to build the Master equation above. This would be a simple procedure if the transitions among each factor did not depend upon each other. However, a key aspect of this sort of DCM is an inherent interdependency among the factors, in which the probability of moving from one state to another—within one factor—depends upon the probability distribution over the states of another factor. For example, the probability that I will move from an asymptomatic to a symptomatic state depends upon the prevalence of infection; namely, the probability that I am infected. These (second-order) dependencies can be expressed as probability transition matrices within each factor that are conditioned upon the levels of another. In what follows, we will go through the four factors describing the second-order dependencies and occasional third-order dependencies. Third-order dependencies mean that the influence of one factor on another depends upon a third factor. These high order dependencies need to be incorporated into the master equation, after it is composed according to Equation 3.

To illustrate how this DCM is parameterised, we will now go through each of the factors in turn. The equations and underlying latent states (i.e., compartments) will be presented as figures, that are followed by intuitive descriptions. Please see the MATLAB code for one or two details that have been omitted for clarity. There are four factors, *location*, *infection*, *symptom*, and *testing*. These factors have four or five levels or states each.

## Location

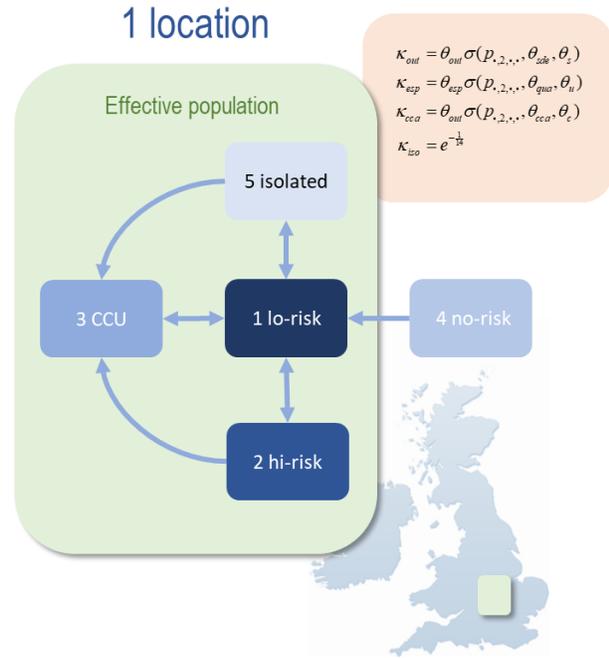

$$P_{\cdot,\cdot,1,\cdot} = \begin{bmatrix} \kappa_{out} & 1 & 1 & 1-\kappa_{ep} & 1-\kappa_{iso} \\ 1-\kappa_{out} & 0 & & & \\ & & 0 & & \\ & & & \kappa_{ep} & \\ & & & & \kappa_{iso} \end{bmatrix} \quad \text{When asymptomatic}$$

$$P_{\cdot,\cdot,2,\cdot} = \begin{bmatrix} 0 & & & & \\ & 0 & & & \\ & & 0 & & \\ & & & 0 & \\ 1 & 1 & 1 & 1 & 1 \end{bmatrix} \quad \text{When symptomatic}$$

$$P_{\cdot,\cdot,3,\cdot} = \begin{bmatrix} 0 & & & & \\ & 0 & & & \\ \kappa_{cca} & \kappa_{cca} & 1 & \kappa_{cca} & \kappa_{cca} \\ & & 0 & & \\ 1-\kappa_{cca} & 1-\kappa_{cca} & 0 & 1-\kappa_{cca} & 1-\kappa_{cca} \end{bmatrix} \quad \text{With respiratory distress}$$

$$P_{\cdot,\cdot,4,\cdot} = \begin{bmatrix} 0 & & & & \\ & 0 & & & \\ & & 0 & & \\ 1 & 1 & 1 & 1 & 1 \\ & & & & 0 \end{bmatrix} \quad \text{When deceased}$$

Stop isolating when asymptomatic and PCR −    Isolate if PCR +        Isolate if infectious (FTTIS efficacy)

$$P_{\cdot,\cdot,1,4} = \begin{bmatrix} \kappa_{out} & 1 & 1 & 1-\kappa_{ep} & 1 \\ 1-\kappa_{out} & 0 & & & \\ & & 0 & & \\ & & & \kappa_{ep} & \\ & & & & 0 \end{bmatrix}, \; P_{\cdot,\cdot,1,3} = \begin{bmatrix} 0 & 1 & 1 & 1-\kappa_{ep} & 1-\kappa_{iso} \\ & 0 & & & \\ & & 0 & & \\ & & & \kappa_{ep} & \\ 1 & & & & \kappa_{iso} \end{bmatrix}, \; P_{\cdot,2,1,11} = \begin{bmatrix} \kappa_{out}(1-\theta_{ft}) & 1 & 1 & 1-\kappa_{ep} & 1-\kappa_{iso} \\ (1-\kappa_{out})(1-\theta_{ft}) & 0 & & & \\ & & 0 & & \\ & & & \kappa_{ep} & \\ \theta_{ft} & & & & \kappa_{iso} \end{bmatrix}$$

**Figure 7:** *location*

The location factor has five states, four of which constitute an *effective* or *affected* population that is a subset of the total or census population. The remainder of the population is assigned to a *no-risk* state. The four states of the effective population include a low and high-risk state. Here, risk refers to the probability of coming into contact with somebody who is infected. Low-risk could be being at home (in an affected area), while high-risk could be being at work or a football match (in an affected area). In addition to these two locations one could be in critical care (requiring mechanical ventilation) or self-isolated. The transition matrices on the left of Figure 7 describe the transitions amongst these location states when *asymptomatic*, *symptomatic*, when severely ill (e.g., acute respiratory distress syndrome, *ARDS*) or when *deceased*. These four states are the levels of the *symptom* factor.

The transition matrices in Figure 7 show that when *asymptomatic*, I have a certain probability of leaving the low-risk (home) location and entering a high-risk (e.g. work) location. When at work, I will inevitably return home in the evening. When in a critical care unit (*CCU*), I will be discharged when and only when *asymptomatic*. In terms of moving from a low-risk state to *isolation*, I will stay in self-isolation for 14 days and return to a low-risk (e.g., domestic setting) after that time. Similarly, if I am *symptomatic*, I will always go into self-*isolation* and when *deceased*, go into a *no-risk* state. When in acute respiratory distress (*ARDS*) I will be taken to critical care (*CCU*) unless I am already there. However, there is also a possibility that the capacity of critical care has been exceeded—and I am not admitted—in which case, I will go into isolation.

The third-order dependencies are described in the lower panel of transition matrices (labelled with green text). For example, I will stop self-*isolating* provided I do not have symptoms and I am PCR negative. Similarly, if I am asymptomatic and test positive, then I will go into *isolation*. The third route to isolation models the efficacy of find test trace isolate and support (FTTIS). This is captured by the probability that I will go into isolation if I am (told by a contact tracer that I am) asymptomatic and infected.

The *no-risk* state (state number four) models a reservoir of the population that has yet to be affected by the epidemic. In this (simplified) description, one can only move from a no-risk to a low-risk state. Effectively, this models the spread of the virus through communities, thereby enlarging the effective population as time goes on. In a full implementation, there is also an efflux from the affected population (i.e., a low-risk state) back into a no-risk state. With this bidirectional exchange—between the effective and no-risk population, the size of the affected population can reach some equilibrium depending upon the relative rates of influx and efflux. We have omitted these effects to avoid visual clutter. In other words, the model in this figure just allows for a progressive increase in the size of the affected population or the number of communities that are exposed to the virus as time proceeds. A simplifying assumption here is that we have used (epidemic) model of a single region—of varying size—as opposed to a (pandemic) model of multiple regions described elsewhere (Friston et al. 2020b).

The format of the equations is reproduced in subsequent figures: $\kappa$ generally refers to a rate constant, namely, the parameters of transition probabilities. The free parameters of the model parameterise these rate constants and are denoted by $\theta$ (the functional form of this parameterisation is provided in the pink boxes). For example, the probability that I will leave home is the product of a baseline probability of going to work times a decreasing sigmoid function of the prevalence of infection. The midpoint of this sigmoid function can be regarded as a soft threshold. In a similar way, the influx of people who have hitherto been in no-risk areas is a decreasing function of prevalence at a lower threshold. It is the disparity between these two (lockdown versus quarantine) thresholds that generates secondary and subsequent waves.

Heuristically, as the virus replicates within the effective population (i.e., the community into which a virus was introduced, such as a city) the lower threshold is crossed and the influx of people into the effective population falls. This can be mediated by travel restrictions, quarantine, or a cordon sanitaire. Within the effective population, prevalence continues to increase until the second threshold is crossed, and within-community distancing is realised through a decreased probability of leaving low-risk locations (i.e., 'stay at home'). Community transmission is attenuated, and the prevalence of infection peaks and then falls as the community acquires a degree of population immunity. As prevalence falls below the lockdown threshold the community unlocks, and social distancing measures are relaxed: for example, the period after the first wave during the summer in the United Kingdom. However, as prevalence falls towards the lower threshold there is an influx from the no-risk locations as travel restrictions are relaxed (e.g., students returning to university) and population immunity is subsequently diluted. The virus then starts to propagate through the extended effective population, engendering a secondary wave.

This succession of between and within-community changes in population mixing and contact rates furnishes a simple model of viral spread throughout the total population, producing a succession of progressively attenuated waves. Note that there is a similar threshold function for entering the critical care state that models a limited capacity. We will see later that the only way to survive severe clinical consequences of infection (e.g., acute respiratory distress syndrome) is to be admitted to critical care. The final parameter determines how long one is in critical care, expressed as a time constant. These time constants can be regarded as expected dwell time or duration that one is in a particular state. This concludes our description of the location model, where transitions among different states depend primarily upon the symptom factor but also have third order dependencies on the infection and testing factors.

## Infection

Figure 8 uses the same format to detail transitions among different states of infection. Here, people start out in a *susceptible* state from which they can get *infected*. From the infected stated there are two routes to an absorbing state (c.f., the removed state of conventional SEIR models). One can either have a mild illness and move straight to a seronegative state (Grifoni et al., 2020; Le Bert et al., 2020; Seo et al., 2020) that may be associated with T-cell mediated humoral immunity (Gallais et al., 2020; Grifoni et al., 2020; Le Bert et al., 2020). Conversely, one can have a more severe illness with viral shedding (van Kampen et al., 2020) and move to an infectious state, and then become seropositive (Ab+) for a period of time (Bao et al., 2020; Houlihan et al., 2020; Ng et al., 2020; Wajnberg et al., 2020). This period of time corresponds with loss of antibodies (Winter and Hegde, 2020), parameterised with the appropriate time constant, causing a transition from seropositive to seronegative immunity (Ab-). The immunity is modelled by precluding a return from seropositive to susceptible, which could be relaxed if necessary.

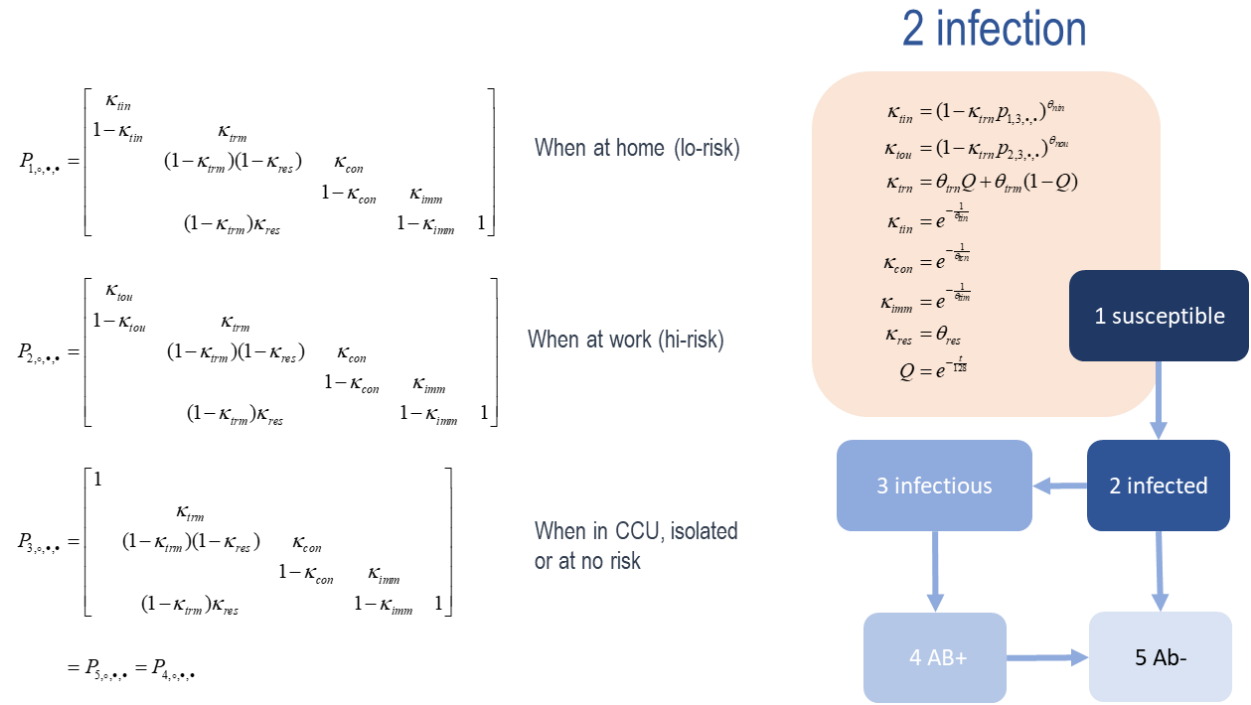

**Figure 8:** *infection*

The rate constants for this factor deal with the probability of becoming *infected*, parameterised in terms of different contact rates. More specifically, the probability of remaining *uninfected* is the probability of avoiding contagion raised to the power of the number of expected contacts per day. In turn, this depends upon whether you are in a low or high-risk situation—or indeed are isolated in CCU or are in a no-risk area. The probability of becoming infected per contact is itself a rate constant times the prevalence of infection in the respective area. In this model, the implicit transmission strength can change over time (either increasing or decreasing). The remaining rate constants are specified in terms of their expected time constants, while the proportion of people who seroconvert is specified by a free parameter. This parameter quantifies the overdispersion or heterogeneity of transmission (Endo et al., 2020; Lloyd-Smith et al., 2005), in the sense that only a certain proportion of people ever become infectious after being infected. This proportion is estimated as a free parameter of the model—analogous to *k* in conventional models. This concludes our description of the infection model that depends on, and only on, the location factor.

## Symptoms

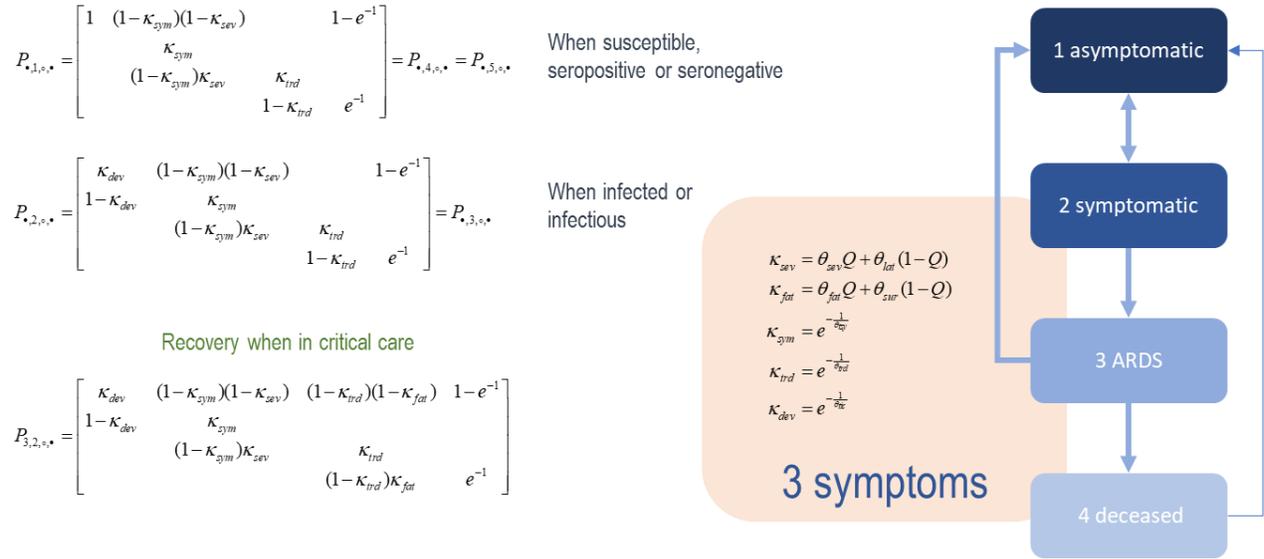

**Figure 9:** *symptoms*

The *symptom* or clinical factor has four states. One can start in an *asymptomatic* state and then, consequent on being *infected*, can—with some probability—become *symptomatic*. A small proportion of *symptomatic* people will progress to potentially fatal respiratory distress and subsequently die, unless they are supported in critical care. The corresponding rate constants are again time-dependent, meaning that the probability of developing *ARDS* can change over different phases of the epidemic (either increase or decrease). Similarly, the probability of dying from *ARDS* can itself change with time, for example as treatments improve. The remaining parameters are specified in terms of the expected dwell times, here in terms of an incubation period, asymptomatic period, and a period of severe disease. Typically, these sum to about 20 days, which is the expected time between becoming infected and death.

## Testing

Testing has four states, starting with having never been tested. One then has a test and *waits* for the results that can either be *positive* or *negative*. Having been tested, one then returns to the *not tested* state. The probability transition matrices are equipped with test sensitivity and specificity parameters, with negative tests when susceptible or no longer infected or infectious (i.e., *seropositive*, or *seronegative*). Conversely, when infected or infectious, there is a high probability of moving to the PCR positive state. Note that when not part of the effective population, one does not bother getting tested.

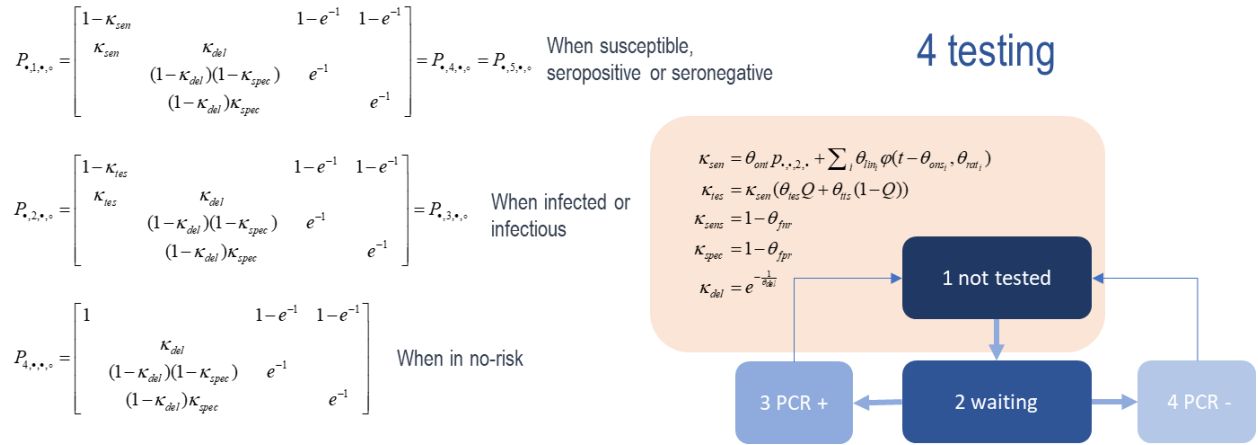

**Figure 10**: *testing*

The probability of submitting to a test depends upon testing capacity that is modelled with a linear mixture of increasing sigmoid function—as various phases of testing are rolled out. In addition, there is a component of test rates that reflects demand in terms of the prevalence of people who are symptomatic. Note further that self-selection bias is parameterised in terms of the relative probability of getting tested if infected, relative to not being infected. This testing bias can change with time (e.g., the relative number of pillar one and pillar two tests in the UK).

This concludes our discussion of the *testing* factor. The transitions among the various states above effectively constitute the prior probabilities that, when equipped with a likelihood model, complete the generative model. The likelihood model maps from the latent states above to observations by taking the expected outcome and adding a random effect.

### The likelihood model

The likelihood (a.k.a. observation) models used in this DCM are based upon simple counting statistics for large numbers. After a square root transform, the data can then be treated as a Gaussian variate with unit variance about the expected number. In practice, because the data are smoothed (which introduces serial dependencies among the data), DCM estimates the random observation effects under fairly informative hyperpriors. In other words, it adjusts the variance or precision of each data modality in proportion to the total number of observations based upon the residual sum of squares. Hyperpriors in this instance specify the prior beliefs about the variance of these residuals, automatically adjusting for the uncertainty in the data, relative to the priors above. In practice, the hyperprior expectation for the variance is set to the log of the summed observations for any kind of data. This automatically adjusts for sparse data with relatively few data points.

**Table 2**: outcomes and their expectations

| Outcome | Description | Likelihood model | Source | Units |
|---|---|---|---|---|
| PCR cases (ONS) | Cases by specimen date (UK total) | $Np_{\bullet,\bullet,\bullet,3}$ | https://coronavirus.data.gov.uk /cases | number /day |
| Daily deaths (ONS) | Deaths within 28 days of positive test by date of death (UK total) | $Np_{\bullet,\bullet,4,\bullet}$ | https://coronavirus.data.gov.uk /deaths | number /day |
| Ventilated patients (ONS) | Patients in mechanical ventilation beds | $N\theta_{cc}p_{3,\bullet,\bullet,\bullet}$ | https://coronavirus.data.gov.uk /healthcare | number |
| PCR tests (ONS) | Number of confirmed positive, negative or void lab-based COVID-19 test results | $Np_{3,\bullet,\bullet,\bullet}$ | https://coronavirus.data.gov.uk /testing | number /day |
| Prevalence (ONS) | Estimate of the number of people testing positive for COVID-19 | $N(p_{\bullet,2,\bullet,\bullet} + p_{\bullet,3,\bullet,\bullet})$ | https://www.ons.gov.uk/peopl epopulationandcommunity/hea lthandsocialcare/conditionsand diseases/datasets/coronavirusc ovid19infectionsurveydata | number |
| Seropositive (GOV) | Number of people testing positive for COVID-19 antibodies | $Np_{\bullet,4,\bullet,\bullet}$ | https://www.ons.gov.uk/peopl epopulationandcommunity/hea lthandsocialcare/conditionsand diseases/datasets/coronavirusc ovid19infectionsurveydata | percent |
| Symptoms (KCL) | Number of people calculated to have COVID symptoms on each day | $N\theta_{sy}(p_{\bullet,\bullet,2,\bullet})$ | https://covid.joinzoe.com/data #levels-over-time | number |
| R-ratio (GOV) | The R number range for the UK | $R_t = \exp(K_t \cdot \theta_{con})$ $K_t = \ln \dfrac{p_{\bullet,2,\bullet,\bullet,t+1}}{p_{\bullet,2,\bullet,\bullet,t}} = \dfrac{\ln(2)}{T_d}$ | https://www.gov.uk/guidance/ the-r-number-in-the-uk#contents | ratio |
| Transport (GOV) | Percentages of the first week in Feb-20 (Cars) | $\theta_{mo1}p_{2,\bullet,\bullet,\bullet}(p_{\bullet,1,\bullet,\bullet})^{\theta_{mo2}}$ $\times 100$ | https://www.gov.uk/governme nt/statistics/transport-use-during-the-coronavirus-covid-19-pandemic | percent |

| Work (Google) | Community Mobility Reports (Workplaces) | $\theta_{wo1} p_{2,\bullet,\bullet,\bullet} (p_{\bullet,1,\bullet,\bullet})^{\theta_{wo2}}$ $\times 100$ | https://www.google.com/covid 19/mobility/ | percent |
|---|---|---|---|---|

*Table 2 lists the outcomes used to invert the dynamic causal model. The predicted outcomes are a straightforward function of the marginal probabilities over various latent states. For example, the percentage of people who are seropositive is the proportion of people in the Ab+ state of the second factor multiplied by one hundred. Some predictions involve exponents. For example, the percentage of people at work is proportional to the probability of being in a high-risk location times the probability that any member of the population has yet to be infected, raised to a power. When this power is very small, this nonlinear term effectively disappears. All these expected outcomes are instantaneous functions of the probability, with the exception of the reproduction ratio that depends on the rate of change of prevalence.*

## Reproduction ratio

The effective reproduction rate is a fundamental epidemiological constant that provides a useful statistic that reflects the exponential growth of the prevalence of infection. There are several ways in which it can be formulated. For our purposes, we can generate an instantaneous reproduction rate directly from the time varying prevalence of infection as follows:

$$R_t = \exp(K_t \cdot \theta_{con})$$

$$K_t = \ln \frac{p_{\bullet,2,\bullet,\bullet_{t+1}}}{p_{\bullet,2,\bullet,\bullet_t}} = \frac{\ln(2)}{T_d}$$

These expressions show that the reproduction rate reflects the growth of (the logarithm of) the proportion of people infected—and the period of being infectious. This number is formally related to the doubling time $T_d$. Note that the reproduction rate is not an estimate in this scheme: it is an outcome that is generated by the latent causes or hidden states inferred by inverting (i.e., fitting) the model to empirical timeseries.

## Software note

The figures in this report can be reproduced using annotated (MATLAB) code available as part of the free and open source academic software SPM (https://www.fil.ion.ucl.ac.uk/spm/), released under the terms of the GNU General Public License version 2 or later. The routines are called by a demonstration script that can be invoked by typing >> DEM_COVID_UK at the MATLAB prompt. At the time of writing, these routines are undergoing software validation in our internal source version control system—that will be released in the next public release of SPM (and via GitHub at https://github.com/spm/). In the interim, please see https://www.fil.ion.ucl.ac.uk/spm/covid-19/.

The data used in this technical report are available for academic research purposes from the sites listed in the fourth column of Table 2.

## Acknowledgements

We gratefully acknowledge comments from Dr Alexander Billig on an earlier draft. This work was undertaken by members of the Wellcome Centre for Human Neuroimaging, UCL Queen Square Institute of Neurology. The Wellcome Centre for Human Neuroimaging is supported by core funding from Wellcome [203147/Z/16/Z]. A.R. is funded by the Australian Research Council (Refs: DE170100128 and DP200100757).

The authors declare no conflicts of interest.